# Are the authors of highly-cited articles also the most productive ones?[1]

*Giovanni Abramo[a], Tindaro Cicero[b], Ciriaco Andrea D'Angelo[a,c]*

[a] Laboratory for Studies of Research and Technology Transfer
at the Institute for System Analysis and Computer Science (IASI-CNR)
National Research Council of Italy

[b] Italian National Agency for the Evaluation of Universities and Research Institutes
(ANVUR)

[c] Department of Management and Engineering
University of Rome "Tor Vergata"

**Abstract**

Ever more frequently, governments have decided to implement policy measures intended to foster and reward excellence in scientific research. This is in fact the intended purpose of national research assessment exercises. These are typically based on the analysis of the quality of the best research products; however a different approach to analysis and intervention is based on the measure of productivity of the individual scientists, meaning the overall impact of their entire scientific production over the period under observation. This work analyzes the convergence of the two approaches, asking if and to what measure the most productive scientists achieve highly-cited articles; or vice versa, what share of highly-cited articles is achieved by scientists that are "non-top" for productivity. To do this we use bibliometric indicators, applied to the 2004-2008 publications authored by academics of Italian universities and indexed in the Web of Science.

# 1. Introduction

In recent decades, the development of the so-called knowledge economy has led many governments to undertake policies and initiatives intended to improve the effectiveness and efficiency of their domestic higher education systems. In particular, there has been increasing implementation of national research assessment exercises, essentially with aims of allocating resources according to merit and stimulating increased levels of research productivity from the funding recipients (Geuna and Martin, 2003; Hicks, 2012). Historically, the conduct of these evaluation exercises has been founded on peer-review methodology, applied to a subset of the overall scientific production that is achieved by the research organizations evaluated. This is the case, for example, of the forthcoming UK's Research Excellence Framework[2] (as well as an earlier series of "RAEs"[3]), which will examine a maximum of three or four of the highest quality works produced by the top scientists selected by the research institutions. A hybrid peer-review/bibliometrics method was adopted in the latest Italian assessment exercise, the 2004-2010 VQR[4], in which universities were required to present, for each of their professors, the best three research works from the period under observation. The formulation of these two assessment exercises, while apparently quite similar, in reality overlies policy objectives that are different, with the first being intended to stimulate excellence among the few, while the second is for upgrading of all. Thus the definition and the measurement of scientific excellence leave space for different formulations and indicators of measure, according to the intended objectives.

In general, scientific excellence of an institution is a multi-dimensional concept (Tijssen, 2003). Its measurement can be conducted through two distinct approaches: from the perspective of the quality of the research products or of the research staff. One example of the first perspective is seen in the so-called Excellence Rate, an indicator used by SCImago in its regular World Report[5], which indicates the percentage of an institution's overall scientific output falling in the set of 10% most-cited papers in the respective scientific fields. Bornmann et al. (2011) have used this indicator to locate centers of excellence at the European level. This perspective in analyzing excellence has also stimulated numerous studies focused on specific sub-fields, both in the hard sciences (for example environmental sciences, Khan and Ho, 2012; or urology, Hennessey et al., 2009) and in social sciences (psychology, in Cho et al., 2012; law, Shapiro, 1991). According to Zitt et al. (2005) "highly-cited articles" is one of the most frequently used indicators for measurement of excellence.

The second perspective instead approaches evaluation from the point of view of the evaluating the research staff of the organizations, meaning that centers of excellence in a field are then recognized for their relative numbers of top scientists in the field (Abramo et al., 2009a); and that two institutions can be compared in terms of productivity of their respective research staff (Abramo et al., 2011a). The literature on research excellence is particularly rich and can be segmented in at least three groups of contributions. The first area in fact concerns the bibliometric indicators proposed for the evaluation of performance in general, and in consequence for the identification of top scientists (Abramo et al., 2013a; Baccini et al., 2012; Bornmann et al., 2007; van Raan,

---

[2] http://www.ref.ac.uk/pubs/, last accessed on June 28, 2013
[3] http://www.rae.ac.uk/, last accessed on June 28, 2013
[4] http://www.anvur.org/sites/anvur-miur/files/bando_vqr_def_07_11.pdf, last accessed on June 28, 2013
[5] http://www.scimagoir.com/pdf/sir_2012_world_report.pdf, last accessed on June 28, 2013

2006; Egghe, 2006; Hirsch, 2005). A second group of works concerns the study of the determinants of performance, particularly the personal and contextual variables that can make a researcher a top scientist (Costas et al., 2012; Abramo et al., 2011b; Abramo et al., 2009b). Finally, a third group of works concerns analysis of the role that top scientists have or should have within their institutional contexts (Prpic, 2011; Silversides, 2010; Ioannidis, 2010; Goodall, 2006).

The question of the definition of excellence comes up again at the level of the individual scientist: is an excellent scientist the one who produces highly-cited articles or the one that has an overall impact on the advancement of knowledge greater than his or her colleagues in the same field? To the best of our knowledge, there seems to have been no exploration of the convergence of these two perspectives of excellence: as defined in terms of individual research products or as defined in terms of the performance of scientists. The current work responds to this gap in the literature by attempting to clarify if the most productive scientists are also those that produce the best articles or if there are meaningful differences between the two perspectives. In fact if they were divergent, then the decision maker would have to be more cautious and precise in choosing how to weigh the concept of excellence, according to the policy objectives being sought.

To provide an exhaustive response, we consider every university researcher active in the hard sciences in Italy. For each individual we measure the scientific productivity, through a bibliometric indicator based on their publications indexed in the Thomson Reuters Web of Science (WoS). The comparison of the value of the productivity indicator measured for all the researcher in a given field then permits identification of the so-called top scientists (TSs), for that field. At the same time, by counting the citations of the publications authored by Italian university professors, and comparing to world publications of the same year and subject category, we identify the highly-cited articles (HCAs). Given this basis, we can advance the following research questions:

i. *Who produces HCAs?* We provide an overall view of who (in terms of TSs and non-TSs) produces HCAs, highlighting potential differences between the fields.
ii. *What is the correlation between research productivity and production of HCAs?* For each researcher, we measure the correlation between their scientific productivity and their production of HCAs.
iii. *What is the distribution of HCAs among Italian scientists?* We analyze the distribution of HCA production per decile and quartile of the researchers, as classified for their scientific productivity.
iv. *Are there differences across academic ranks?* We attempt to understand if the answer to the first research question is different for full, associate and assistant professors.

**2. Methodology**

**2.1 Identifying excellence among research results**

In the hard sciences, the prevalent form of codification of research output is publication in peer reviewed journals. For this reason we assume that excellent results are observable in the form of excellent publications. The excellence of a publication is demonstrated by its placement in the high extremes of the scale of value shared by the international scientific community of the specific discipline. The value of a publication is understood as its impact on scientific advancement, and as proxy of this impact bibliometricians adopt the number of citations for the publication itself. Thus to identify excellent publications, we refer to lists ranking the publications by the number of citations they receive, for all publications indexed in WoS or Scopus of the same year and subject category. In the current work we define excellent or "highly-cited articles" (HCAs) as those that place in the top five percent of the citation ranking list for WoS-indexed publications of the same year and subject category.

**2.2 Identifying excellence among scientists**

We can define excellent scientists as those that place above a certain rank in a performance ranking list formed of comparable units, i.e. scientists working in the same field. The main performance indicator of any production system is productivity, so we can define the top 5% of scientists ranked for productivity as "excellent".

To assess scientific productivity of individual researchers in the hard sciences (Abramo et al., 2013a) we consider the outcome, meaning the impact of their research activities, over the five year period from 2004 to 2008. As a proxy of outcome we adopt the number of citations (observed at 31/12/2011), for each academic's publications indexed in WoS, as is common practice for the hard sciences. Because the intensity of publications varies by field (Garfield, 1979; Moed et al., 1985; Butler, 2007; Abramo and D'Angelo, 2007), in order to avoid distortions in productivity rankings (Abramo et al., 2008) we then compare professors within the same field. In the Italian academic system, each professor is classified in one and only one research field. There are a total of 370 such fields (scientific disciplinary sectors, or SDSs[6]), grouped into 14 disciplines (university disciplinary areas, or UDAs). When measuring labor productivity, if there are differences in the production factors available to each scientist then there should be normalization for them. Unfortunately, relevant data at the individual level are not available in Italy. Another issue is that it is very possible that professors belonging to a particular scientific field will also publish outside that field. Because citation behavior varies by field, we thus standardize the citations for each publication with respect to the average of the distribution of citations for all the cited Italian publications of the same year and the same WoS subject category[7]. Furthermore, research projects frequently involve a team of scientists, which shows in co-authorship of publications. In this case we account for the fractional contribution of the scientists to output by equating it to the reciprocal of the number of co-authors. The contributions of the individual co-authors to the achievement of the publication are not necessarily equal, and in some fields the authors signal the different contributions through their order in the byline. For life sciences, in order to avoid distortions (Abramo et al., 2013b), we give different weights to each co-author according to his/her position in the byline and the character of the co-authorship (intra-mural or extra-mural)[8]. Thus in formula, the proxy for yearly

---

[6] The complete list is accessible at http://attiministeriali.miur.it/UserFiles/115.htm, last accessed on June 28, 2013.
[7] As shown by Abramo et al. (2012), the average of the distribution of citations received for all cited publications of the same year and subject category is the most reliable scaling factor for Italian data.
[8] If first and last authors belong to the same university, 40% of citations are attributed to each of them; the remaining 20% are divided among all other authors. If the first two and last two authors belong to different organizations, 30% of citations are attributed to first and last authors; 15% of citations are attributed to second and last author but one; the remaining 10% are divided among all others.

productivity of a single researcher, *FSS*, is:

$$FSS = \frac{1}{t} \cdot \sum_{i=1}^{N} \frac{c_i}{\bar{c}_i} f_i$$

[1]

Where:
*t* = number of years of work of the researcher in the period of observation;
*N* = number of publications of the researcher in the period of observation;
$c_i$ = citations received by publication *i*;
$\bar{c}_i$ = average citations received by all cited publications of the same year and subject category of publication *i*;
$f_i$ = fractional contribution of the researcher to publication *i*; for the life sciences, different weights are given to each co-author according to their order in the byline and the character of the co-authorship (intra-mural or extra-mural).

The period of observation of production is 2004-2008, and citations are counted as of 31/12/2011. Data on faculty of each university and their SDS classification are extracted from the database on Italian university personnel, maintained by the Ministry of University and Research. The bibliometric dataset used to measure *p* is extracted from the Italian Observatory of Public Research (ORP), a database developed and maintained by the authors and derived under license from the WoS. Beginning from the raw data of the WoS, and applying a complex algorithm for reconciliation of the author's affiliation and disambiguation of the true identity of the authors, each publication (article, article review and conference proceeding[9]) is attributed to the university scientist or scientists that produced it (D'Angelo et al., 2011). Thanks to this algorithm, which is unique in the world to the best of our knowledge, we can produce Italian rankings of research productivity at the individual level. Thus the productivity of each scientist is calculated in each SDS and can be expressed on a percentile scale of 0-100 (worst to best) for comparison with the productivity of all Italian colleagues of the same SDS. We define a top scientist, TS, as a researcher whose performance rank by FSS, in their SDS, is above the 95th percentile.

For the WoS-indexed publications to serve as a more robust proxy of overall output of a researcher, the field of observation is limited to those SDSs where at least 50% of member scientists produced at least one publication in the period 2004-2008. There are 182 such SDSs[10]. We further exclude those researchers active less than three years over the observation period. The dataset for the analysis thus consists of over 35,000 professors and almost 150,000 publications (Table 1).

---

[9] The dataset excludes document types not recognizable as true research products (editorial material, meeting abstracts, news items, etc.), as well as publications for which Thomson Reuters does not provide world percentile citations (letters, etc.).
[10] The complete list is accessible at http://www.disp.uniroma2.it/laboratoriortt/testi/Indicators/ssd3.html

| UDA | N. of SDSs | Professors | Publications* |
|---|---|---|---|
| Mathematics and computer science | 9 | 3,193 | 11,696 |
| Physics | 8 | 2,593 | 20,907 |
| Chemistry | 12 | 3,253 | 23,658 |
| Earth sciences | 12 | 1,280 | 4,281 |
| Biology | 19 | 5,173 | 26,717 |
| Medicine | 47 | 10,942 | 49,231 |
| Agricultural and veterinary sciences | 27 | 2,772 | 9,192 |
| Civil engineering | 7 | 1,317 | 3,007 |
| Industrial and information engineering | 41 | 4,900 | 20,029 |
| Total | 182 | 35,423 | 148,367** |

*Table 1: Number of SDSs, researchers, and publications in each UDA*
*\* The figure refers to publications authored by at least one professor pertaining to the UDA.*
*\*\* Total is less than the sum of the column data due to double counts of publications co-authored by researchers pertaining to more than one UDA.*

## 3. Results

*Who produces highly-cited articles?*

The bibliometric analysis of the dataset articles shows that 5.2% of the total meet the definition of "highly cited", thus representing a value slightly higher than expected, if one presumes that the ratio of excellent to total works is independent of scale. But who produces these HCAs? How many of these HCAs are produced by TSs? The data in Table 2 indicate that 32.9% are produced exclusively by TSs, while 36.5% are produced exclusively by non-TSs. The remainder (25.4%) are obviously coauthored by both TSs and their "non-top" colleagues. TSs authored or coauthored 58.3% of HCAs; non-TS contribution is 67.1%.

| UDAs | N. HCAs* | Of which authored by TSs | Of which authored by TSs only | Of which authored by non-TSs | Of which authored by non-TSs only |
|---|---|---|---|---|---|
| Civil engineering | 126 (4.2%) | 66.7% | 54.0% | 46.0% | 24.7% |
| Mathematics and computer science | 581 (5.0%) | 55.6% | 44.2% | 55.8% | 44.4% |
| Earth sciences | 236 (5.5%) | 51.3% | 40.7% | 59.3% | 41.8% |
| Industrial and information engineering | 988 (4.9%) | 52.7% | 37.0% | 63.0% | 42.8% |
| Biology | 1,218 (4.6%) | 52.8% | 36.9% | 63.1% | 34.4% |
| Agricultural and veterinary sciences | 391 (4.3%) | 55.5% | 32.7% | 67.3% | 34.8% |
| Chemistry | 1,178 (5.0%) | 53.7% | 30.4% | 69.6% | 37.1% |
| Medicine | 2,766 (5.6%) | 58.9% | 30.4% | 69.6% | 31.0% |
| Physics | 1,181 (5.6%) | 42.9% | 27.3% | 72.7% | 41.9% |
| Total | 7,685** (5.2%) | 58.3% | 32.9% | 67.1% | 41.7% |

*Table 2: Highly cited articles (ratio to total output in brackets) and their authors in each UDA*
*\* The figure refers to HCAs authored by at least one researcher pertaining to the UDA.*
*\*\* Total is less than the sum of the column data due to double counts of publications co-authored by researchers pertaining to more than one UDA.*

The analyses for the single UDAs reveal significant differences: in Mathematics and computer science the share of HCAs is almost equally distributed between the two groups: 55.6% are produced by TSs, 55.8% by the remaining non-TSs. However in Civil engineering the TSs coauthored 66.7% of the HCAs compared to 46.0% from the non- TSs. The data show this is clearly quite a unique UDA: it is the one with the lowest absolute number of HCAs (126), and at the same time an important share of these is produced exclusively by TSs (54.0%). On the opposite front, the data for the Physics UDA are striking: here only 27.3% of HCAs are authored exclusively by TSs. In effect, research in this UDA, particularly in the fields of particle and high-energy physics, is often conducted through so-called "grand experiments", where the results, typically of high scientific impact, are accredited to all the research staff of the partner organizations. If we exclude the two extremes, represented by Civil engineering on the one hand and Physics on the other, the other UDAs distribute in two clusters: the first group, composed of Biology, Chemistry, Medicine and Agricultural and veterinary sciences, shows percentages of HCAs produced by TSs alone that fall between 30.4% and 36.9%; the second cluster, composed of Mathematics and computer science, Industrial and information engineering and Earth sciences, shows a percentage of HCAs produced by TSs alone that falls between 37.0% and 44.2%. Mathematics and computer science stands out because the non-TSs alone in this UDA produce 44.4% of the total HCAs. Similar levels from the non-TSs are also reached in Industrial and information engineering (42.8%) and Earth sciences (41.8%), as well as obviously for Physics (41.9%).

In Table 3 we see that out of a total of 1,844 TSs in the hard sciences, 81.5% produce at least one HCA. However this average percentage disguises quite heterogeneous behavior at the level of the individual UDAs. In fact the Physics and Chemistry UDAs have a percentage well above the average. In Physics, only 14 TSs out of 133 are without HCAs, while in Chemistry this figure drops to only 12 TSs out of 168 in total. In six UDAs (Civil engineering, Industrial and information engineering, Agricultural and veterinary sciences, Earth science, Biology and Mathematics), the percentage of TSs that produce HCAs drops below the 80% threshold.

| UDAs | N. of TSs | Of which producing HCAs | N. of non-TSs | Of which producing HCAs |
|---|---|---|---|---|
| Chemistry | 168 | 156 (92.9) | 3,085 | 792 (25.7) |
| Physics | 133 | 119 (89.5) | 2,460 | 925 (37.6) |
| Medicine | 564 | 474 (84.0) | 10,378 | 1,947 (18.8) |
| Mathematics and computer science | 163 | 130 (79.8) | 3,030 | 325 (10.7) |
| Biology | 267 | 208 (77.9) | 4,906 | 768 (15.7) |
| Industrial and information engineering | 261 | 202 (77.4) | 4,639 | 620 (13.4) |
| Earth sciences | 69 | 53 (76.8) | 1,211 | 155 (12.8) |
| Agricultural and veterinary sciences | 150 | 112 (74.7) | 2,622 | 321 (12.2) |
| Civil engineering | 69 | 49 (71.0) | 1,248 | 63 (5.0) |
| Total | 1,844 | 1,503 (81.5) | 33,579 | 5,916 (17.6) |

*Table 3: Number of TSs (top scientists) and non-TSs that produce HCAs (percentage values in brackets)*

However, as emerged from the previous analysis, the contribution to top literature at the world level is also provided by scientists that are not classified as national top per productivity. This phenomenon comes up particularly in the Physics and Chemistry UDAs, where over 37.6% and 25.7% of non-TSs produce exceptional results at the world level. In Civil engineering, in contrast, the contribution of non-TSs is quite marginal, at approximately 5%.

From this first analysis, it would seem there is a certain link between the two perspectives in analysis of excellence: individuals considered as TSs also produce a certain share of HCAs at the international level. However this pattern is not constant in all UDAs, as might have been expected: in some disciplines a very relevant share of HCAs is produced by non-TSs. To examine this aspect further, in the next section we analyze the correlation between a researcher's scientific productivity and their HCA production.

*What is the correlation between scientific productivity and production of HCA?*

To respond to this research question we proceed by codifying HCA production as a dummy variable, with value 1 when the researcher produces at least one HCA, otherwise 0. This transformation is suggested by the fact that the variable is notably concentrated and asymmetric: approximately 79% of scientists do not produce any HCAs and 98.5% author less than five. Rather than applying a linear correlation coefficient it appears more reasonable to resort to a point-biserial correlation coefficient (Tate, 1954), which permits analysis of the relationship between a continuous variable (scientific productivity) and a binary variable (HCA production or non-production). For each SDS, the coefficient of correlation (*r*) between FSS and production of HCA is, in formula [2].

$$r = \frac{\overline{FSS_p} - \overline{FSS_q}}{SD} * \sqrt{pq}$$

[2]

Where:
$\overline{FSS_p}$: average value of FSS of individuals that produce at least one HCA.
$\overline{FSS_q}$: average value of FSS of individuals that produce no HCAs.
SD: standard deviation of the entire distribution of FSS
p: number of researchers with variable dummy = 1
q: number of researchers with variable dummy = 0

From the analysis, 77% of SDSs have a correlation coefficient between 0.386 and 0.653, meaning in the interval average ± one standard deviation, a percentage much greater than what would be expected under a Gaussian distribution (68.27%). The correlation values concentrate around the average (0.52), with a quite symmetric distribution. Given these observations, we view it as reasonable to consider a value below 0.3 as weak correlation and values over 0.6 as strong correlation. The correlation results are summarized in Table 4. For each UDA, the table presents the statistics describing the correlation between productivity and production of HCAs at the SDS level.

We observe that in the Chemistry UDA, correlation varies over a quite contained interval (0.417 to 0.594). This confirms the observations of the preceding analysis, in which it emerged that in this UDA a relevant share of HCAs are produced by non-TSs. The Mathematics and computer sciences UDA is the one with the greatest share of SDSs with a strong correlation. In Mathematical logic (MAT/01) this correlation exceeds the threshold of 0.90. This pattern appears again in the UDAs Civil engineering and Industrial and information engineering: in Environmental and health engineering (ICAR/03), the value of maximum correlation is over 0.827 and in Nuclear measurement tools (ING-IND/20) it exceeds 0.95.

In the opposite situation of weak correlation values, we observe the Physics UDA, with 25% of the SDSs having a value under 0.3. Many of the SDSs in this UDA are characterized by papers with many authors (at times over a thousand), thus we expect a high number of weak correlations, since there will be numerous academics that author HCAs while there are relatively few TSs. In effect a specific analysis shows a very low value of average correlation (0.391) for the SDSs in the Physics UDA, compared to other UDAs.

| UDAs | N. of SDSs* | Correlation | | | |
|---|---|---|---|---|---|
| | | Weak (≤0.3) | Strong (≥0.6) | Min | Max |
| Mathematics and computer science | 9 | 0 (0%) | 5 (56%) | 0.454 (MAT/02) | 0.905 (MAT/01) |
| Physics | 8 | 2 (25%) | 1 (13%) | 0.148 (FIS/04) | 0.717 (FIS/06) |
| Chemistry | 11 | 0 (0%) | 0 (0%) | 0.417 (CHIM/02) | 0.594 (CHIM/04) |
| Earth sciences | 12 | 0 (0%) | 2 (17%) | 0.412 (GEO/09) | 0.664 (GEO/10) |
| Biology | 19 | 0 (0%) | 2 (11%) | 0.358 (BIO/11) | 0.667 (BIO/08) |
| Medicine | 47 | 2 (4%) | 3 (6%) | 0.208 (MED/32) | 0.688 (MED/45) |
| Agricultural and veterinary sciences | 27 | 1 (4%) | 8 (30%) | 0.176 (AGR/09) | 0.736 (AGR/04) |
| Civil engineering | 7 | 2 (29%) | 3 (43%) | 0.153 (ICAR/05) | 0.827 (ICAR/03) |
| Industrial and information engineering | 38 | 3 (8%) | 12 (32%) | -0.039 (ING-IND/04) | 0.950 (ING-IND/20) |

*Table 4: Statistics for point-biserial correlation between FSS and production of HCAs, for the SDSs of each UDA*
* CHIM/05 and ING-IND/28 to 30 are excluded due to low numbers of observations. See
http://www.disp.uniroma2.it/laboratoriortt/testi/Indicators/ssd3.html for full names of SDSs.

*What is the distribution of HCAs among Italian scientists?*

To respond to this question we first subdivide the distribution of FSS for each UDA in deciles of equal range, the first decile being the top 10% of researchers per FSS in a given UDA. For each interval we calculate the number of HCAs produced by the scientists of that decile. Table 5 presents the example of the Mathematics and computer sciences UDA, showing the contributing of each decile of scientists to production of HCAs. An HCA co-authored by researchers ranked in two or more deciles is assigned to all deciles involved. Table 5 thus presents the frequencies of HCA authorship. In this case for the 581 HCAs produced in Mathematics and computer science, there are 783 authorships. At the decile level we can further see that in the case of this UDA the top 10% of researchers (10th decile class for productivity) produce 63.9% (500/783) of HCA authorships.

| Decile class for FSS | Frequency | Relative Frequency | Cumulative |
|---|---|---|---|
| 1 [0-10] | 0 | 0.0 | 0.0 |
| 2 ]10-20] | 0 | 0.0 | 0.0 |
| 3 ]20-30] | 0 | 0.0 | 0.0 |
| 4 ]30-40] | 3 | 0.4 | 0.4 |
| 5 ]40-50] | 5 | 0.6 | 1.0 |
| 6 ]50-60] | 16 | 2.0 | 3.1 |
| 7 ]60-70] | 28 | 3.6 | 6.6 |
| 8 ]70-80] | 72 | 9.2 | 15.8 |
| 9 ]80-90] | 159 | 20.3 | 36.1 |
| 10 ]90-100] | 500 | 63.9 | 100.0 |

*Table 5: Distribution of frequencies of HCA authorship in Mathematics and computer sciences, per decile class for FSS*

Table 6 presents the relative frequency values for all nine UDAs. The Chemistry UDA stands out as having the least skewness, with researchers in the first productivity decile even authoring some HCAs (although only 0.1% of the total of the HCA authorships). Also notable is Physics, where researchers with productivity below the median value (from first to fifth decile) are responsible for 14.7% of the HCA authorships in the UDA. Otherwise this percentage never rises above 6% in all the other UDAs, reaching a maximum of 5.7% in Biology and Medicine. HCA authorships from the top 10% of professors is decidedly high in Civil engineering (76.9%) and Mathematics and computer sciences (63.9%), while in these same two disciplines the researchers who place below the national median for productivity are responsible for only around 1% of the HCA authorships.

| UDAs | Deciles | | | | | | | | | |
|---|---|---|---|---|---|---|---|---|---|---|
| | 1 | 2 | 3 | 4 | 5 | 6 | 7 | 8 | 9 | 10 |
| Mathematics and computer science | 0.0 | 0.0 | 0.0 | 0.4 | 0.6 | 2.0 | 3.6 | 9.2 | 20.3 | 63.9 |
| Physics | 0.0 | 1.0 | 2.8 | 4.8 | 6.1 | 6.7 | 9.2 | 11.3 | 30.1 | 28.1 |
| Chemistry | 0.1 | 0.4 | 1.3 | 1.7 | 2.2 | 4.7 | 7.1 | 10.2 | 17.5 | 54.8 |
| Earth sciences | 0.0 | 0.6 | 0.3 | 0.6 | 2.0 | 4.9 | 7.2 | 11.5 | 17.9 | 55.0 |
| Biology | 0.0 | 0.4 | 0.9 | 1.7 | 2.7 | 4.7 | 6.0 | 11.5 | 16.9 | 55.2 |
| Medicine | 0.0 | 0.3 | 0.8 | 1.8 | 2.7 | 4.1 | 6.5 | 10.1 | 18.2 | 55.4 |
| Agricultural and veterinary sciences | 0.0 | 0.1 | 0.5 | 1.6 | 2.8 | 3.8 | 5.3 | 11.8 | 17.6 | 56.4 |
| Civil engineering | 0.0 | 0.0 | 0.0 | 1.1 | 0.0 | 1.1 | 3.8 | 3.8 | 13.2 | 76.9 |
| Industrial and information engineering | 0.0 | 0.1 | 0.4 | 0.7 | 1.6 | 5.1 | 6.4 | 10.4 | 18.4 | 56.9 |
| Total | 0.0 | 0.5 | 1.3 | 2.3 | 3.2 | 4.9 | 7.0 | 10.6 | 20.9 | 49.4 |

*Table 6: Relative frequency (%) of authorship of HCAs by decile class for FSS, per UDA*

Classification of the researchers in quartiles for FSS makes the differences among UDAs still more apparent (Table 7). We note that the top quarter of authors contribute 90% of HCA authorships in Civil Engineering and Mathematics, while the minimum value from the top quarter (63.6%) is registered in Physics. For all the other UDAs, the incidence of HCA authorship from the scientists in the top quartile varies between 78% and 82% of total.

|                                      | Quartiles |      |      |      |
|--------------------------------------|-----------|------|------|------|
| UDAs                                 | I         | II   | III  | IV   |
| Mathematics and computer science     | 0.0       | 1.0  | 9.3  | 89.7 |
| Physics                              | 2.3       | 12.4 | 21.8 | 63.6 |
| Chemistry                            | 1.3       | 4.4  | 16.2 | 78.1 |
| Earth sciences                       | 0.6       | 2.9  | 17.6 | 79.0 |
| Biology                              | 0.9       | 4.8  | 15.6 | 78.6 |
| Medicine                             | 0.6       | 5.1  | 15.3 | 79.0 |
| Agricultural and veterinary sciences | 0.3       | 4.8  | 14.7 | 80.3 |
| Civil engineering                    | 0.0       | 1.1  | 7.1  | 91.8 |
| Industrial and information engineering | 0.4     | 2.5  | 15.3 | 81.9 |
| Total                                | 1.1       | 6.2  | 16.7 | 76.0 |

*Table 7: Relative frequency of HCA authorship per quartile by FSS, per UDA*

*Are there differences among academic ranks?*

The objective of the final analysis is to reveal potential differences between academic ranks in production of HCAs, in terms of the production from the TSs and non-TSs of each rank. For this we apply a statistical instrument commonly used for case-control studies in epidemiology, involving calculation of the so-called "odds ratio" (OR). For our purposes, we stratify the TSs by academic rank (assistant, associate e full) and then evaluate for the presence of differences in the answer to the above specific research question. For each UDA and academic rank we thus subdivide the observations in two sets:
- the "cases", meaning the professors who author HCAs;
- the "controls", meaning the professors that did not produce any HCAs.

The dataset is thus now represented by a 2x2 contingency table. Table 8 presents the specific example of the case of the assistant professors in the Medicine UDA.

|         | Cases | Controls |
|---------|-------|----------|
| TS      | 51    | 12       |
| Non TS  | 571   | 3,963    |
| Total   | 622   | 3,975    |

*Table 8: Contingency table for case-control study and calculation of the odds ratio for assistant professors in Medicine*

This table is used to calculate the odds ratio, dividing the odds for the cases by the odds for the controls (Bland and Altman, 2000). The odds are the ratio of the probability that the event of interest occurs (that is, an HCA is authored by a TS) to the probability that it does not. As seen in the example of Table 8, the odds for the cases are calculated as the ratio of 51 to 571 (0.089), while the odds for the controls are the ratio of 12 to 3,963 (0.003). The quotient of these two quantities is the odds ratio, in this case equal to 29.5. This signifies that the probability of producing HCAs for the TSs is roughly 30 times greater than for those who are non-TSs. Table 9 presents the summary results of these calculations per UDA: to carry out the comparison between ranks, we pair the data concerning assistant and full professors (here omitting the associate professors). Although our analysis is not based on a sample but on the overall population of Italian academics in the hard sciences, we also provide the 95% confidence intervals.

|  | Assistant professors | | | Full professors | | |
|---|---|---|---|---|---|---|
|  | OR | 95% interval | | OR | 95% interval | |
| UDAs |  | Lower | Upper |  | Lower | Upper |
| Mathematics and computer science | 28.3 | 9.9 | 91.0 | 32.1 | 17.7 | 61.4 |
| Physics | 9.8 | 2.8 | 53.0 | 11.3 | 5.1 | 29.5 |
| Chemistry | 7.4 | 2.0 | 32.9 | 42.1 | 18.4 | 118.2 |
| Earth sciences | 20.2 | 5.4 | 91.2 | 20.2 | 7.6 | 62.3 |
| Biology | 8.0 | 3.3 | 19.5 | 18.6 | 10.9 | 24.6 |
| Medicine | 29.5 | 15.4 | 61.2 | 16.8 | 10.1 | 18.4 |
| Agricultural and veterinary sciences | 12.2 | 4.9 | 31.2 | 20.2 | 11.3 | 37.6 |
| Civil engineering | 10.6 | 1.5 | 58.3 | 84.1 | 31.7 | 242.7 |
| Industrial and information engineering | 18.4 | 7.9 | 46.2 | 26.3 | 16.2 | 44.0 |
| Test di homogeneity (H – M) | Chi2=10.96 | *P-value*=0.204 | | Chi2=26.41 | *P-value*=0.001 | |

*Table 9: Calculation of odds ratio (OR) and 95% confidence interval for professor ranks, by UDA*

For assistant professors, the lower limit of the confidence interval is always greater than one, thus indicating a positive relationship between being a TS and the production of HCAs. The Mantel-Haenszel homogeneity test shows that there is not strong inhomogeneity between the UDAs (*P-value* = 0.204), in spite of the fact that the ORs clearly vary, between values:
- less than 10, in Physics (9.8), Chemistry (7.4) and Biology (8.0);
- greater than 20, in Earth sciences (20.2), Mathematics and computer sciences (28.3) and Medicine (OR=29.5).

In contrast, the analysis for full professors produces some surprises. For this role, the OR values are almost always higher than those registered for assistant professors: in Civil engineering the value is actually eight times higher (84.1 vs. 10.6). The homogeneity test shows the presence of significant differentiation between the UDAs (*P-value* = 0.001).

The probability that a non-TS produces HCA is certainly greater for assistant professors than for full professors, in all the UDAs except for Earth sciences (where ORs are equal to 20.2 for both assistant and full professors), and this is especially the case for Medicine. In this last UDA, the probability of producing HCA for full professors TSs is 16.8 times greater than for non-TSs, while for assistant professors this relationship is almost doubled (29.5).

**Conclusions**

In recent years, governments of various nations have implemented research assessment exercises with greater frequency, typically based on the analysis of the quality of the best research products of the organizations evaluated. A different perspective of evaluation for organizations is to examine the productivity of their respective research staffs. In this work we have analyzed the potential convergence of the two perspectives, attempting to clarify if the most productive scientists are also the ones that produce highly-cited articles.

The analysis of scientific production of Italian academics in the hard sciences, for the period 2004-2008, shows that 58.3% of highly-cited articles are produced by the top-productive scientists (32.9% by TSs only), while 67.1% are produced by non-TSs (41.7% by non-TSs only). At the same time it emerges that almost 20% of TSs did not produce any HCAs over the five years examined; in the same period 17.6% of the non-TSs authored at least one HCA.

Thus there is a moderate correlation between the phenomenon of being a top-productive scientist and the probability of having produced HCAs. This result is confirmed by an analysis of the frequency of co-authorship by decile for productivity: about half of the co-authorships of HCAs is due to scientists in the first decile for productivity.

There is further an undoubted differentiation between the disciplines: the point bi-serial correlation between productivity and HCA production is highest in Mathematics and computer science and lowest in Physics.

Again in Mathematics, 63.9% of HCA authorships is due to scientists in the first decile; in Civil engineering this statistic even rises to 76.9%, while in Physics it registers at 28.1%. Thus we can say that particularly in Physics the convergence of the two perspectives of excellence is weak, logically because in this discipline research is often conducted through so called "grand experiments", where the results are accredited to all the research staff in the organizations that took part.

Other than the disciplinary differentiations indicated, the study also evidences differences between academic ranks: the probability of non-TSs producing HCAs is greatest among assistant professors, in all the UDAS, with the sole exception of Medicine.

For the evaluator interested in incentivizing excellence in research, the results of this work thus seem to suggest that the choice of the most appropriate indicator or combination is critical and need to be aligned with the objectives to be achieved.